\documentclass[10pt,twocolumn]{article}
\usepackage[dvips]{graphicx}
\usepackage{myusenix}

\usepackage{amsmath}
\begin{document}

\renewcommand\dbltopfraction{.95}
\renewcommand\textfraction{.05}

\title{CUP: Controlled Update Propagation in Peer-to-Peer Networks}

\author{
Mema Roussopoulos \hspace{0.75cm} Mary Baker\\ 
{\em Department of Computer Science} \\
{\em Stanford University}\\
{\em Stanford, California, 94305}\\
\\
\normalsize \{mema, mgbaker\}@cs.stanford.edu \\
 http://mosquitonet.stanford.edu/} 

\date{February 1, 2002}

\maketitle 

\begin{abstract}

\small Recently the problem of indexing and locating content in
peer-to-peer networks has received much attention.  Previous work
suggests caching index entries at intermediate nodes that lie on the
paths taken by search queries, but until now there has been little
focus on how to maintain these intermediate caches.  This paper
proposes CUP, a new comprehensive architecture for Controlled Update
Propagation in peer-to-peer networks.  CUP asynchronously builds
caches of index entries while answering search queries.  It then
propagates updates of index entries to maintain these caches.  Under
unfavorable conditions, when compared with standard caching based on
expiration times, CUP reduces the average miss latency by as much as a
factor of three.  Under favorable conditions, CUP can reduce the
average miss latency by more than a factor of ten.
 
CUP refreshes intermediate caches, reduces query latency, and reduces
network load by coalescing bursts of queries for the same item.  CUP
controls and confines propagation to updates whose cost is likely to
be recovered by subsequent queries.  CUP gives peer-to-peer nodes the
flexibility to use their own incentive-based policies to determine
when to receive and when to propagate updates.  Finally, the small
propagation overhead incurred by CUP is more than compensated for by
its savings in cache misses.

\end{abstract}

\section{Introduction}

Peer-to-peer systems are self-organizing distributed systems where
participating nodes both provide and receive services from each other
in a cooperative effort to prevent any one node or set of nodes from
being overloaded.  Peer-to-peer systems have recently gained much
attention, primarily because of the great number of features they
offer applications that are built on top of them.  These features
include: scalability, availability, fault tolerance, decentralized
administration, and anonymity.

Along with these features has come an array of technical challenges.
In particular, over the past year, there has been much focus on the
fundamental indexing and routing problem inherent in all peer-to-peer
systems: Given the name of an object of interest, how do you locate
the object within the peer-to-peer network in a well-defined,
structured manner that avoids flooding the network \cite{ratnasamy01a,
rowstron01b, stoica01, zhao01}?

As a performance enhancement, the designers of these systems suggest
caching index entries with expiration times at intermediate nodes that
lie on the path taken by a search query.  Intermediate caches are
desirable because they balance query load for an item across multiple
nodes, reduce latency, and alleviate hot spots. However, little
attention has been given to how to maintain these intermediate caches.
This problem is interesting because the peer-to-peer model assumes the
index will change constantly.  This constant change stems from several
factors: peer nodes continuously join and leave the network, content
is continuously added to and deleted from the network, and replicas of
existing content are continuously added to alleviate bandwidth
congestion at nodes holding the content.

In this paper we propose a new comprehensive architecture for
Controlled Update Propagation (CUP) in peer-to-peer networks that
asynchronously builds caches of index entries while answering search
queries.  It then propagates updates of index entries to maintain
these caches.  The basic idea is that every node in the peer-to-peer
network maintains two logical channels per neighbor: a query channel
and an update channel.  The query channel is used to forward search
queries for items of interest to the neighbor that is closest to the
authority node for those items.  The update channel is used to forward
query responses (first-time updates) asynchronously to a neighbor and
to update index entries that are cached at the neighbor.

\begin{figure}[tb]
\centerline{\includegraphics[height=7cm]{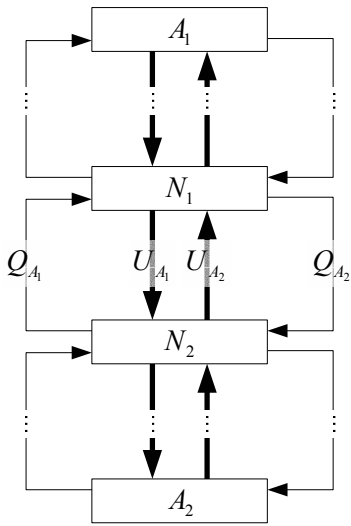}}
\caption[]{\small This figure shows how CUP works.  \(A_1\) and
\(A_2\) are authority nodes.  \(Q_{A_1}\), \(Q_{A_2}\), \(U_{A_1}\),
and \(U_{A_2}\) are the four logical channels between nodes \(N_1\)
and \(N_2\).  A query  arriving at node \(N_2\) for an item for which
\(A_1\) is the authority is pushed onto query channel \(Q_{A_1}\) to
\(N_1\).  If \(N_1\) has a cached entry for the item, it returns it through
\(U_{A_1}\).  Otherwise, it forwards the query  towards \(A_1\).
Any update originating from \(A_1\) flows
downstream to \(N_1\) which may forward it onto \(N_2\) through
\(U_{A_1}\).  The analogous process holds for queries
at \(N_1\) for items for which \(A_2\) is the authority.}
\label{fig:logicalChannels}
\end{figure}

Queries for an item travel ``up'' the query channels of nodes along
the path toward the authority node for that item.  Updates travel
``down'' the update channels along the reverse path taken by a query.
Figure~\ref{fig:logicalChannels} shows this process.

The advantages of the query channel are twofold.  First, if a node
receives two or more queries for an item for which it does not have a
fresh response, the node pushes only one instance of the query for
that item up its query channel.  
This approach can have
significant savings in traffic, because bursts of requests for an item
are coalesced into a single request.  Second, using a single query
channel solves the ``open connection'' problem suffered by some
peer-to-peer systems.  Each time a query arrives at a node which does
not have a cached response, the node opens one or more connections to
neighboring nodes and must maintain those connections open until the
response returns through them.  The asynchronous nature of the query
channel relieves nodes from having to maintain many open connections
since all responses return through the update channel.  Through simple
bookkeeping (setting an interest bit) the node registers the interest
of its neighbors so it knows which of its neighbors to push the query
response to when the answer arrives.

The cascaded propagation of updates from authority nodes down the
reverse paths of search queries has many advantages.  First, updates
extend the lifetime of cached entries allowing intermediate nodes to
continue serving queries from their caches without having to push
queries up their channels explicitly.  It has been shown that up to
fifty percent of content hits at caches are instances where the
content is valid but stale and therefore cannot be used to serve
queries without first being re-validated \cite{cohen01b}.  These occurrences are
called \emph{freshness misses}. Second, a node that
proactively pushes updates to interested neighbors reduces its load of
queries generated by those neighbors.  The cost of pushing the update
down is recovered  by the first query for the same item
following the update.
Third, the further down an update gets pushed, the shorter the
distance subsequent queries need to travel to reach a fresh cached
answer.  As a result, query response latency is reduced.  Finally,
updates can help prevent errors.  For example, an update to invalidate
an index entry prevents a node from answering queries using the entry
before it expires.

In CUP, nodes decide individually when to receive updates. A node only
receives updates for an item if the node has registered interest in
that item.  Furthermore, each node uses its own incentive-based policy to
determine when to cut off its incoming supply of updates for an
item.  This way the propagation of updates is controlled and does not
flood the network.

Similarly, nodes decide individually when to propagate updates to
interested neighbors.  This is useful because a node may not always be
able or willing to forward updates to interested neighbors. In fact, a
node's ability or willingness to propagate updates may vary with its
workload.  CUP addresses this by introducing an adaptive mechanism
each node uses to regulate the rate of updates it propagates
downstream.  A salient feature of CUP is that even if a node's 
capacity to push updates becomes zero, nodes dependent on the
node for updates fall back with no overhead to the case
of standard caching with expiration.

When compared with standard caching, under unfavorable conditions, CUP
reduces the average miss latency by as much as a factor of three.
Under favorable conditions, CUP reduces the average miss latency by
more than a factor of ten.  CUP overhead is more than compensated for
by its savings in cache misses.  In fact, the ``investment'' return
per update pushed in saved misses grows substantially with
increasing network size and query rates.  The cost of saved misses can
be one to two orders of magnitude higher than the cost of updates
pushed.

We demonstrate that the performance of CUP depends highly on the
policy a node uses to cut off its incoming updates. We find that the
cut-off policy should adapt to the node's query workload and we
present probabilistic and log based methods for doing so.  Finally, we
show that CUP continues to outperform standard caching even when
update propagation is reduced by either node capacity or network
conditions.

The rest of the paper is organized as follows:
Section~\ref{Architecture} describes in detail the design of the CUP
architecture.  Section~\ref{Evaluation} describes the cost model we use to
evaluate CUP and presents experimental evidence of the benefits of CUP.
Section~\ref{RelatedWork} discusses related work and
Section~\ref{Conclusions} concludes the paper.

\section{CUP Architecture Design}
\label{Architecture}

First, we provide some background terminology we use throughout the
paper and very briefly describe how peer-to-peer networks for which
CUP is appropriate perform their indexing and lookup operations.
Then we describe the components of the CUP protocol.

\subsection{Background} 

The following terms will be useful for the remainder of the paper:

\emph{Node}: This is a node in the peer-to-peer network.  Each node
periodically exchanges ``keep-alive'' messages with its neighbors to
confirm their existence and to trigger recovery mechanisms should one
of the neighbors fail.  
Every node also maintains two logical channels (connections) for each neighbor:
the query channel and the update channel.  The query channel is used
by the node to push queries to its neighbor.  The update channel is
used by the node to push updates that are of interest to the neighbor.

\emph{Global Index}: The most important operation in a peer-to-peer
network is that of locating content.  As in \cite{ratnasamy01a} we
assume a hashing scheme that maps keys (names of content files or
keywords) onto a virtual coordinate space using a uniform hash
function that evenly distributes the keys to the space.  The
coordinate space serves as a global index that stores index entries
which are \emph{(key, value)} pairs.  The value in an index entry is a
pointer (typically an IP address) to the location of a replica that
stores the content file associated with the entry's key.  There can be
several index entries for the same key, one for each replica of the
content.

\emph{Authority Node}: Each node N in the peer-to-peer system is
dynamically allocated a subspace of the coordinate space (i.e., a
partition of the global index) and all index entries mapped into its
subspace are owned by N. We refer to N as the authority node of these
entries.  \emph{Replicas} of content whose key corresponds to an
authority node N send birth messages to N to announce they are willing
to serve the content.  Depending on the application supported,
replicas might periodically send refresh messages to indicate they are
still serving a piece of content.  They might also send deletion
messages that explicitly indicate they are no longer serving the
content.  These deletion messages trigger the authority node to delete
the corresponding index entry from its local index directory.

\emph{Search Query}: A search query posted at a node N is a request to
locate a replica for key K.  The response to such a search query is a
set of index entries that point to replicas that serve the content
associated with K.

\emph{Query Path for Key K}: This is the path a search query for key
\emph{K} takes.  Each hop on the query path is in the direction of
the authority node that owns \emph{K}.  If an intermediate node on this
path has fresh entries cached, the path ends at the intermediate
node; otherwise the path ends at the authority node.

\emph{Reverse Query Path for Key K}: This path is the reverse of the
query path defined above.

\emph{Local index directory}: This is the subset of  global index
entries owned by a node.

\emph{Cached index entries}: This is the set of index entries cached
by a node N in the process of passing up queries and propagating down
updates for keys for which N is not the authority.  The set of
cached index entries and the local index directory are disjoint sets.

\emph{Lifetime of index entries}: We assume that each index entry
cached at a node has associated with it a lifetime and a timestamp
indicating the time at which the lifetime was set.  When the
difference between the current time and the timestamp is greater than
the lifetime field, the entry has expired and cannot be used to answer
queries.  An index entry is considered fresh until it expires.

\subsection{How Routing Works}

We assume that anytime a node issues a query for key \emph{K}, the
query will be routed along a well-defined structured path with a
bounded number of hops from the querying node to the authority node
for \emph{K}.  The routing mechanism is designed so that each node on
the path hashes \emph{K} using the same hash function to deterministically choose
which of its neighbors will serve as the next hop.
Examples of peer-to-peer systems that provide this type of structured
location and routing mechanism include content-addressable networks
(CANs) \cite{ratnasamy01a}, Chord \cite{stoica01}, Pastry
\cite{rowstron01b} and Tapestry \cite{zhao01}.  CUP can be used in the
context of any of these systems.

\subsection{Node Bookkeeping}

At each node, index entries are grouped together by key.  For each key
K, the node stores a flag that indicates whether the node is waiting
to receive an update for K for the first time and an interest bit
vector.  Each bit in the vector corresponds to a neighbor and is set
or clear depending on whether that neighbor is or is not interested in
receiving updates for K.

Each node tracks the popularity or request frequency of each non-local
key K for which it receives queries.  The popularity measure for a key
K can be the number of queries for K a node receives between arrivals
of consecutive updates for K or a rate of queries of a larger moving
window.  On an update arrival for K, a node uses its popularity
measure to re-evaluate whether it is beneficial to continue caching and
receiving updates for K.  We elaborate on this cut-off decision in
Section~\ref{CutOffPolicies}.

Node bookkeeping in CUP involves no network overhead. With increasing
CPU speeds and memory sizes, this bookkeeping is negligible when we
consider the reduction in query latency achieved.

\subsection{Update Types}

We classify updates into four categories: first-time updates, deletes,
refreshes, and appends.  Deletes, refreshes, and
appends originate from the replicas of a piece of content and are
directed toward the authority node that owns the index entries for
that content.  

First-time updates are query responses that travel down the reverse
query path.

Deletes are directives to remove a cached index entry.  Deletes can be
triggered by two events: 
1) a replica sends a message indicating it no
longer serves a piece of content to the authority node that owns the
index entry pointing to that replica. 
2) the authority node notices a
replica has stopped sending ``keep-alive'' messages and assumes the
replica has failed.  In either case, the authority node deletes the
corresponding index entry from its local index directory and
propagates the delete to interested neighbors.

Refreshes are keep-alive messages that extend the lifetimes of index
entries.  
Refreshes that arrive at a cache do not result in errors as deletes
do, but help prevent freshness misses.

Finally, appends are directives to add index entries for new replicas
of content.  These updates help alleviate the demand for content from the
existing set of replicas since they add to the number of replicas from
which clients can download content.

\subsection{Handling Queries}
\label{HowQueryingWorks}

Upon receipt of a query for a key \emph{K}, there are three basic
cases to consider.  In each of the cases, the node updates its
popularity measure for \emph{K}.  The node also sets the appropriate
bit in the interest bit vector for \emph{K} if the query originates
from a neighbor.  Otherwise, if the query is from a local client, the
node maintains the connection until it can return a fresh answer to the
client.
To simplify the protocol description we use the phrase
``push the query'' to indicate that a node pushes a query upstream
toward the authority node.  We use the phrase ``push the update'' to
indicate that a node pushes an update downstream in the direction of
the reverse query path.

\textbf{Case 1: Fresh Entries for key K are cached.} 
The node uses its cached entries for \emph{K} to push the response
as a first-time update to the querying neighbor or local client.

\textbf{Case 2: Key K is not in cache.}  The node adds \emph{K} to its
cache and marks it with a \emph{Pending-First-Update} flag.  The
purpose of the \emph{Pending-First-Update} flag is to coalesce bursts
of queries for the same key into one query.  A subsequent query for
\emph{K} from a neighbor or a local client will save the node from
pushing another instance of the query for \emph{K}.

\textbf{Case 3: All cached entries for key K have expired.}  The node
must obtain the fresh index entries for \emph{K}.  If the
\emph{Pending-First-Update} flag is set, the node does not need to
push the query; otherwise, the node sets the flag and pushes the
query.  

\subsection{Handling Updates}
\label{UpdateArrivals}

A key feature of CUP is that a node does not forward an update for
\emph{K} to its neighbors unless those neighbors have registered
interest in \emph{K}.  Therefore, with some light bookkeeping, we
prevent unwanted updates from wasting network bandwidth.  

Upon receipt of an update for key \emph{K} there are three cases to
consider.

\textbf{Case 1: Pending-First-Update flag is set.}  This means that
the update is a first-time update carrying a set of index entries in
response to a query.  The node stores the index entries in its cache,
clears the \emph{Pending-First-Update} flag, and pushes the update to
neighbors whose interest bits are set and to local client connections
open at the node.

\textbf{Case 2: Pending-First-Update flag is clear.}  If all the
interest bits for \emph{K} are clear, the node decides whether it
wants to continue receiving updates for \emph{K}.  The node bases its
decision on \emph{K}'s popularity measure.  Each node uses its own
policy for deciding whether the popularity of a key is high enough to
warrant receiving further updates for it.  If the node decides
\emph{K}'s popularity is too low, it pushes a \emph{Clear-Bit} control
message to the neighbor from whom it received the update.  The
\emph{Clear-Bit} message indicates that the neighbor's interest bit
for this node should be cleared.  Otherwise, if the popularity is high
or some interest bits are set, the node applies the update to its
cache and pushes the update to the neighbors whose bits are set.

Note that a greedy or selfish node can choose not to push updates for
a key K to interested neighbors.  This forces downstream nodes to fall
back to standard caching for K.  However, by choosing to cut off
downstream propagation, a node runs the risk of receiving subsequent
queries from its neighbors.  Handling each of these queries is twice
the cost of propagating an update downward because the node has to
receive the query from the downstream neighbor and then push the
response as an update.  Therefore, although each node is free to stop
pushing updates at any time it is in its best interest to push updates
for which there are interested neighbors.

\textbf{Case 3: Incoming update has expired.}  This could occur when
the network path has long delays and the update does not arrive in
time.  The node does not apply the update and does not push it downstream.

\subsection{Handling Clear-Bit Messages}
\label{Clear-Bit-Messages}

A \emph{Clear-Bit} control message is pushed by a node to indicate to
its neighbor that it is no longer interested in updates for a
particular key from that neighbor.

When a node receives a \emph{Clear-Bit} message for key K, it clears
the interest bit for the neighbor from which the message was sent.  If
the node's popularity measure for K is low and all of its interest
bits are clear, the node also pushes a \emph{Clear-Bit} message for K.
This propagation of \emph{Clear-Bit} messages toward the authority
node for K continues until a node is reached where the popularity of K
is high or where at least one interest bit is set.

\emph{Clear-Bit} messages can be piggy-backed onto queries or updates
intended for the neighbor, or if there are no pending queries or
updates, they can be pushed separately.  

\subsection{Adaptive Control of Update Push}
\label{ControllingUpdateArrivals}

Ideally every node would propagate all updates to interested neighbors
to save itself from having to handle future downstream misses.
However, from time to time, nodes are likely to be limited in their
capacity to push updates downstream.  Therefore, we introduce an
adaptive control mechanism that a node uses to regulate the rate of
pushed updates.

We assume each node N has a capacity U for pushing updates that varies
with N's workload, network bandwidth, and/or network connectivity.  N
divides U among its outgoing update channels such that each channel
gets a share that is proportional to the length of its queue.  This
allocation maintains the queues roughly equally sized.  The queues are
guaranteed to be bounded by the expiration times of the entries in the
queues.  So even if a node has its update channels completely shut
down for a long period, all entries will expire and be removed from
the queues.

Under a limited capacity and while updates are waiting in the queues,
each node can re-order the updates in its
outgoing update channels by pushing ahead updates that are likely to
have greater impact on query latency reduction, on query accuracy, or
on the load balancing of content demand across replicas.  During the
re-ordering any expired updates are eliminated.

The strategy for re-ordering depends on the application.  For example,
in an application where query latency and accuracy are of the most
importance, one can push updates in the following order: first-time
updates, deletes, refreshes, and appends. In an application subject to
flash crowds that query for a particular item, appends might be given
higher priority over the other updates.  This would help distribute
the load faster across the entire set of replicas.

A node can also re-order refreshes and appends so that entries that
are closer to expiring are given higher priority.  Such entries are
more likely to cause freshness misses which in turn trigger a new query search.
So it is advantageous to try to catch this in time by pushing these first.

\subsection{Node Arrivals and Departures}

The peer-to-peer model assumes that participating nodes will
continuously join and leave the network.  CUP must be able to handle
both node arrivals and departures seamlessly.  

\textbf{Arrivals.} 
When a new node N enters the peer-to-peer 
network, it becomes the authority node for a portion of the index
entries owned by an existing node M.  N, M, and all surrounding
affected nodes (old neighbors of M) update the bookkeeping structures
they maintain for indexing and routing purposes.  To support CUP, the
issues at hand are updating the interest bit vectors of the affected
nodes and deciding what to do with the index entries stored at M.

Depending on the indexing mechanism used, the cardinality of the bit
vectors of the affected nodes may change.  That is, bit vectors may
expand or contract as some nodes may now have more or fewer neighbors
than before N's arrival.  Since all nodes already need to track who
their neighbors are as part of the routing mechanism, updating the
cardinality of the interest bit vectors to reflect N's arrival is
straightforward.  For example, nodes that now have both N and M as
neighbors have to increase their bit vectors by one element to include
N.  The affected nodes also need to modify the mappings from bit ID to
neighbor IP address in their bit vectors.  For example, if a node that
previously had M as its neighbor now has N as its neighbor, the node
must make the bit ID that pointed to M now point to N.

To deal with its stored index entries, M could simply not hand over
any of its entries to N.  This would cause entries at some of M's
previous neighbors to expire and subsequent queries from those nodes
will restart update propagations from N.  Alternatively, M could give
a copy of its stored index entries to N.  Both N and M would then go
through each entry and patch its bit vector.  This way nodes that
previously depended on M for updates of particular keys could continue
to receive updates from either M or N but not both.

\textbf{Departures.}  Node departures can be either graceful (planned)
or ungraceful (due to sudden failure of a node).  In either case the
index mechanism in place dictates that a neighboring node M take over
the departing node N's portion of the global index.  To support CUP,
the interest bit vectors of all affected nodes must be patched to
reflect N's departure.

If N leaves gracefully, N can choose not to hand over to M its index
entries.  Any entries at surrounding nodes that were dependent on N to
be updated will simply expire and subsequent queries will restart
update propagations.  Again, alternatively N may give M its set of
entries.  M must then merge its own set of index entries with N's, by
eliminating duplicate entries and patching the interest bit vectors as
necessary.  If N's departure is not planned, there can be no hand over
of entries and all entries in the affected neighboring nodes will
expire as in standard caching.

Note that the transfer of entries can be coincided with the transfer
of information that is already occurring as part of the routing mechanism
in the peer-to-peer network, and therefore does not add extra network
overhead.  Also the bit vector patching is a local operation that affects
only each individual node.  Therefore even in cases where a node's
neighborhood changes often, the effect on the overall performance of
CUP is limited to that node's neighborhood (see section 3.7).

\subsection{CUP Query/Update Trees}

\begin{figure}[tb]
\centerline{\includegraphics[width=7cm, height=7cm]{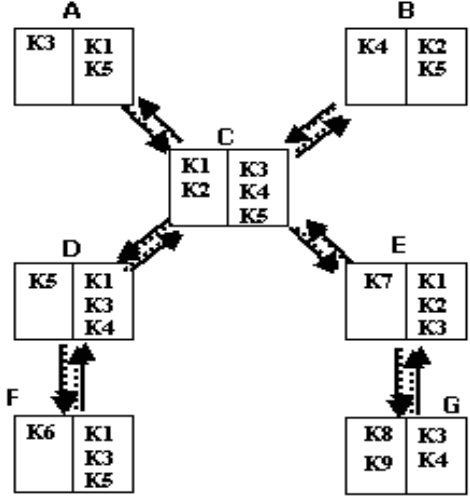}}
\caption[]{\small CUP Trees}
\label{fig:CUPTrees}
\end{figure}

Figure~\ref{fig:CUPTrees} shows a snapshot of CUP in progress in a
network of seven nodes.  The left hand side of each node shows the set
of keys for which the node is the authority.  The right hand side
shows the set of keys for which the node has cached index entries as a
result of handling queries.  For example, node A owns
K3 and has cached entries for K1 and K5.

For each key, the authority node that owns the key is the root of a
CUP tree.  The branches of the CUP tree are formed by the paths
traveled by queries from other nodes in the network.  For example, one
path in the tree rooted at A is \{F, D, C, A\}.

Updates originate at the root (authority node) of a CUP tree and
travel downstream to interested nodes.  Queries travel upstream toward
the root.

\section{Evaluation}
\label{Evaluation}

The goal of CUP is to extend the benefits of standard caching based on
expiration times.  There are two key performance questions to address.
First, by how much does CUP reduce the average cost per query?
Second, how much overhead does CUP incur in providing this reduction?

We first present the cost model based on economic incentive used by
each node to determine when to cut off the propagation of updates for a
particular key.  We give a simple analysis of how the cost per query
is reduced (or eliminated) through CUP.  We then describe our
experimental results comparing the performance of CUP with that of
standard caching.

\subsection{Cost Model}
\label{CostModel}

Consider an authority node A that owns key K and consider the tree
generated by issuing a query for K from every node in the peer-to-peer
network.  The resulting tree, rooted at A, is the \emph{Virtual Query
Spanning Tree} for K, V(A,K), and contains all possible query paths
for K.  The \emph{Real Query Tree} for K, R(A,K) is a subtree of
V(A,K) also rooted at A and contains all paths generated by real
queries.  The exact structure of R(A,K) depends on the actual workload
of queries for K.  The entire workload of queries for all keys results
in a collection of criss-crossing Real Query Trees with overlapping
branches.

We first consider the case of standard caching at the intermediate
nodes along the query path for key K.  Consider a node N that is at
distance D from A in V(A,K).  We define the cost per query for K at N
as the number of hops in the peer-to-peer network that must be
traversed to return an answer to N.  When a query for K is posted at N
for the first time, it travels toward A looking for the response.  If
none of the nodes between N and A have a fresh response cached, the
cost of the query at N is \(2 D\): D hops to reach A and D hops for
the response to travel down the reverse query path as a first-time
update.  If there is a node on the query path with a fresh answer
cached, the cost is less than \(2 D \).  Subsequent queries for K at N
that occur within the lifetime of the entries now cached at N have a
cost of zero.  As a result, caching at intermediate nodes has the
benefits of balancing query load for K across multiple nodes and
lowering average latency per query.

We can gauge the performance of CUP by calculating the percentage of
updates CUP propagates that are ``justified''.  We precisely define
what a justified update is below, but simply put, an update is
justified if it recovers the overhead it incurs, i.e., if its cost is
recovered by a subsequent query.  An unjustified update is therefore
overhead that is not recovered (wasted).  Updates for popular keys are
likely to be justified more often than updates for less popular keys.

A refresh update is justified if a query arrives sometime
between the previous expiration of the cached entry and the new
expiration time supplied by the refresh update.  An append
update is justified if at least one query arrives between
the time the append is performed and the time of its expiration.

A first-time update is always justified because it carries a query's
response toward the node that originally issues the query.  A deletion
update is considered justified if at least one query arrives between
the time the deletion is performed and the expiration time of the
entry to be deleted. 

For each update, let \(T\) be the critical time interval described
above during which a query must arrive in order for the update to be
justified.  (For first-time updates \(T = \infty\)).  Consider a node
N at distance D from A  in R(A,K).  An update propagated down to N is justified
if at least one query Q is posted within \(T\) time units at any of the
nodes of the virtual subtree V(N,K).  Note that an update  is justified
if Q arrives at the virtual tree V(N,K), \emph{not} the real query
tree R(N,K) because Q can be posted anywhere in V(N,K).

Given the distribution of query arrivals at each node in the tree
V(N,K), we can find the probability that the update at N is justified
by calculating the probability that a query will arrive at some node
in V(N,K).  Assume that queries for K arrive at each node \(N_i\) in
V(N,K) according to a Poisson process with parameter \(\lambda_i\).
Then it follows that queries for K arrive at V(N,K) according to a
Poisson process with parameter \(\Lambda\) equal to the sum of all
\(\lambda's\).  Therefore, the probability that a query for K will
arrive within \(T\) time units is \(1 - e^{-\Lambda T}\) and  equals
the probability that the update pushed to N is justified.
The closer to the authority N is, the higher the \(\Lambda\)
and thus the higher the probability for an update pushed to N to be justified.
For \(\Lambda=1\) query arrival per second and  \(T=6\) 
seconds, the probability that an update arriving at N is justified 
is 99 percent.

The benefit of a justified update goes beyond recovery of its cost.
For each hop an update is pushed down, exactly one hop is saved since
without the propagation, a subsequent query arriving within \(T\) time
units would have to travel one hop up and one hop down.  This halves
the number of hops traveled which reduces query response latency, and
at the same time provides enough benefit margin for more
aggressive CUP strategies.
For example, a more aggressive strategy would be to push some updates
even if they are not justified.  As long as the number of justified
updates is at least fifty percent the total number of updates pushed,
the overall update overhead is completely recovered.  If the
percentage of justified updates is less than fifty percent, then the
overhead will not be fully recovered but query latency will be further
reduced.  Therefore, if network load is not the prime concern, an
``all-out'' push strategy achieves minimum latency.

\subsection{Experiments}
\label{Experiments}

One of the challenges in evaluating this work is the unavailability of
real data traces of completely decentralized peer-to-peer networks
such as those assumed by CUP.  The reason for this is that such
systems \cite{ratnasamy01a, rowstron01b, stoica01, zhao01} are not yet
in widespread use to make collecting traces feasible.
Therefore, in the evaluation of CUP we choose simulation parameters
that range from unfavorable to favorable conditions for CUP in order
to show the spectrum of performance and how it compares to standard
caching under the same conditions.  For example, low query rates do
not favor CUP because updates are less likely to be justified since
there may not be enough subsequent queries to recover the cost of the
updates.  On the other hand, queries for keys that become suddenly hot
not only justify the propagation overhead, but also enjoy a significant
reduction in latency. 

For our experiments, we simulated a two-dimensional ``bare-bones''
content-addressable network (CAN) \cite{ratnasamy01a} using the
Stanford Narses simulator \cite{maniatis01}.  The simulation takes as
input the number of nodes in the overlay peer-to-peer network, the
number of keys owned per node, the distribution of queries for keys,
the distribution of query inter-arrival times, the number of replicas
per key, and the lifetime of replicas in the system.  
We ran experiments for n = \(2^k\) nodes where k ranged from 3 to 12.
Simulation time was 22000 seconds, with 3000 seconds of querying time.
We present results for experiments with replica lifetime of 300
seconds to reflect the dynamic nature of peer-to-peer networks where
nodes might only serve content for short periods of time.  For all
experiments, refreshes of index entries occur at expiration.  Query
arrivals were generated according to a Poisson process.
Nodes were randomly selected to post the queries.  

We present five experiments.  First we compare the performance and
overhead of CUP against standard caching where CUP propagates updates
without any concern for whether the updates are justified.  In this
experiment, we vary the level in the CUP tree to which updates are
propagated.  We use this experiment to establish the level that
provides the maximum benefit and then use the performance results at
this level as a benchmark for comparison in later experiments.
Second, we compare the effect on CUP performance of different
incentive-based cut-off policies and compare the performance of these
policies to that of the benchmark.  Third, using the best cut-off
policy of the previous experiment, we study how CUP performs as we
vary the size of the network.  Fourth, we study the effect on
performance of increasing the number of replicas corresponding to a
key.  Finally, we study the effect of limiting the outgoing update
capacities of nodes.  

\subsection{Varying the CUP Push Level}

In this set of experiments we compare standard caching with a version
of CUP that propagates updates down the Real Query Tree of a key
regardless of whether or not the updates are justified.  We use this
information to determine a maximal performance baseline.  We determine
the reduction in misses achieved by CUP and the overhead CUP incurs to
achieve this reduction.  We define \emph{miss cost} as the total
number of hops incurred by all misses, i.e.  freshness and first-time
misses. We define the CUP overhead as the total number of hops
traveled by all updates sent downstream plus the total number of hops
traveled by all clear-bit messages upstream. (We assume clear-bit
messages are not piggybacked onto updates.  This somewhat inflates the
overhead measure.)  We define \emph{total cost} as the sum of the
\emph{miss cost} and any overhead hops incurred.  Note that in
standard caching, the \emph{total cost} is equal to the \emph{miss
cost}.

Figures~\ref{fig:PushLevQ1,10} and \ref{fig:PushLevQ100,1000} plot
CUP's total cost and miss cost versus the push level for a network of
\(2^{10}\) nodes.  A push level of \(p\) means that updates are
propagated to all nodes that have queried for the key and that are at
most \(p\) hops from the authority node. A push level of \(0\)
corresponds to the case of standard caching, since all updates from
the authority node (the root of the CUP tree) are immediately
squelched and not forwarded on.  For this set of experiments, query
arrivals were generated according to a Poisson process with average
rate \(\lambda\) of 1, 10, 100, and 1000 queries per second at the
network.

The figures show that as the push level increases, CUP significantly
reduces the miss cost when compared with standard caching and does
so with little overhead as shown by the displacement of each pair 
of curves. 

In Figure~\ref{fig:PushLevQ1,10}, for $\lambda = 1$ query per second,
the total cost incurred by CUP decreases and reaches a minimum at
around push level 20, after which it slightly increases. This
turning point is the level beyond which the overhead cost of updates is
not recoverable.
For $\lambda = 10$ queries per second, a similar turning
point occurs at around push level 25.  In
Figure~\ref{fig:PushLevQ100,1000} the minimum total cost occurs at
push level 25 and tapers off for both $\lambda =100$ and $\lambda
=1000$ queries per second. For low query arrival rates, the turning
point occurs at lower push levels.  For example, for $\lambda = 0.01$
queries per second, the turning point occurs at push level 15.  These
results show that there is no specific optimal push level at which CUP
achieves the minimum total cost across all workloads.  If there were,
then the simplest strategy for CUP would be to have updates be
propagated to that optimal push level.  In fact, we have found that in
addition to the query workload, the optimal push level is affected by
the number of nodes in the network and the rate at which updates are
generated, both of which change dynamically.

In the absence of an optimal push level, each node needs a policy for
determining when to stop receiving updates.  We next examine some
cut-off policies.

\begin{figure}[tb]
\centerline{\includegraphics[width=9cm]{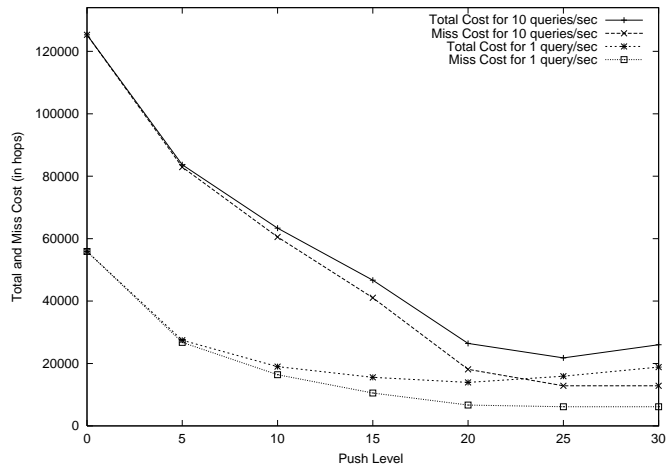}}
\caption{\small Total cost and miss cost versus push level. }
\label{fig:PushLevQ1,10}
\end{figure}

\begin{figure}[tb]
\centerline{\includegraphics[width=9cm]{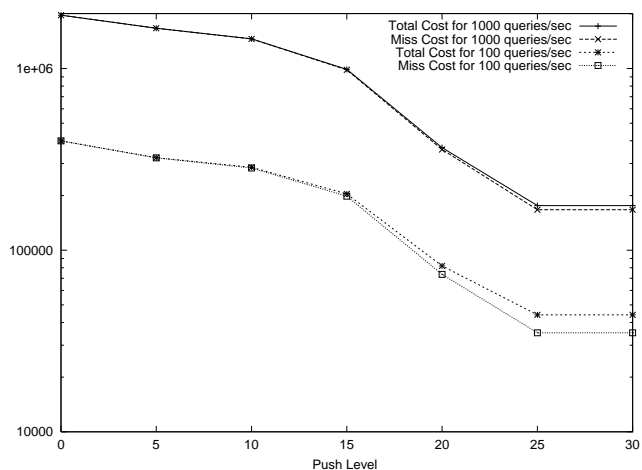}}
\caption{\small Total cost and miss cost versus push level. The
y-axis is log scale.}
\label{fig:PushLevQ100,1000}
\end{figure}

\subsection{Varying the Cut-Off Policies}
\label{CutOffPolicies}

On receiving an update for a key, each node determines 
whether or not there is incentive to continue receiving updates
or to cut off updates by pushing up a clear-bit message.  We base the
incentive on the \emph{popularity} of the key at the node. 
The more popular a key is, the more incentive there is to receive updates
for that key.  For a key K, the popularity is the number of queries a
node has received for K since the last update for K arrived at the
node.  

We examine two types of thresholds against which to test a key's
popularity when making the cut-off decision: probability-based and
log-based.

A probability-based threshold uses the distance of a node N from the
authority node A to approximate the probability that an update pushed
to N is justified.  Per our cost model of section 3.2, the further N
is from A, the less likely an update at N will be justified.  
We examine two such thresholds, a linear one and a logarithmic one.  With a
linear threshold, if an update for key K arrives at a node at distance
$D$ and the node has received at least $\alpha D$ queries for K since
the last update, then K is considered popular and the node continues
to receive updates for K.  Otherwise, the node cuts off its intake of
updates for K by pushing up a clear-bit message.  The logarithmic
popularity threshold is similar.  A key K is popular if the node has
received $\alpha \lg(D)$ queries since the last update.  The
logarithmic threshold is more lenient than the linear  in that it
increases at a slower rate as we move away from the root.

A log-based threshold is one that is based on the recent history of
the last \emph{n} update arrivals at the node. 
If within \emph{n} updates, the node has not received
any queries, then the key is not popular and the node pushes up a clear-bit message.  A specific example of a
log-based policy is the second-chance policy.  In this policy, $n=3$.
When an update arrives, if no queries have arrived since the last
update, the policy gives the key a ``second chance'' and does not push
a clear-bit message immediately.  If at the next update arrival the
node has still not received any queries for K, then it pushes a
clear-bit message.  The philosophy behind this policy is that pushing
these two updates down from the parent node costs two hops.  If a
query arrives in the interval between these two updates,
then it will recover the cost of
pushing them down, since a query miss would incur
two hops, one up and one down.

\begin{table*}
\caption[]{\small Total cost for varying cut-off policies.}
\label{tab:policies2}
\begin{center}
\begin{tabular}{|l|r|r|r|r|} \hline
\textsf{\textbf{Policy}} & \textsf{\textbf{1 q/s Total Cost}} 
& \textsf{\textbf{10 q/s Total Cost}} & \textsf{\textbf{100 q/s Total Cost}}
 & \textsf{\textbf{1000 q/s Total Cost}} \\\hline
Standard Caching & 55905 (1.00) & 125288 (1.00) & 399669 (1.00) & 1967452 (1.00) \\ \hline
Linear, $\alpha = 0.25$ & 61778 (1.11)& 65668 (0.52)& 61521 (0.15)& 188772 (0.10)\\ \hline
Linear, $\alpha = 0.10$ & 39707 (0.71)& 44259 (0.35)& 56170 (0.14)& 184907 (0.09)\\ \hline
Linear, $\alpha = 0.01$ & 28234 (0.51)& 37396 (0.30)& 53613 (0.13)& 182198 (0.09)\\ \hline
Linear, $\alpha = 0.001$ & 28234 (0.51)& 37396 (0.30)& 53613 (0.13)& 182198 (0.09)\\ \hline
Logarithmic, $\alpha = 0.5$ & 30415 (0.54)& 39325 (0.31) & 54816 (0.14)& 183476 (0.09)\\ \hline
Logarithmic, $\alpha = 0.25$ & 28165 (0.50)& 37392 (0.30) & 53613 (0.13) & 182183 (0.09)\\ \hline
Logarithmic, $\alpha = 0.10$ & 28165 (0.50)& 37392 (0.30) & 53613 (0.13) & 182183 (0.09)\\ \hline
Logarithmic, $\alpha = 0.01$ & 28165 (0.50)& 37392 (0.30) & 53613 (0.13) & 182183 (0.09)\\ \hline
Second-chance & 15183 (0.27) & 24886 (0.20) & 46385 (0.12) & 177755 (0.09)\\ \hline
Optimal push level & 13916 (0.25) & 21805 (0.17) & 44062 (0.11) & 175914 (0.09) \\ \hline
\end{tabular}
\end{center}
\end{table*}

Table~\ref{tab:policies2} compares the total cost of standard caching
with that of the linear and logarithmic policies for various $\alpha$
values, the second chance policy, and that of the optimal push level.
The experiment is for a network of \(2^{10}\) nodes and \(\lambda\)
rates of 1, 10, 100 and 1000 queries per second.  In each table entry,
the first number is the total cost and the number in the parentheses
is the total cost normalized by the total cost for standard caching.

For the lower query rates, the performance of the linear and the
logarithmic policies is greatly affected by the choice of parameter
$\alpha$.  In some cases, the total cost of the linear policy exceeds
that of standard caching.  For both the linear and logarithmic
policies, the total cost decreases with $\alpha$ until it reaches a
minimum that cannot be reduced any further. For both $\lambda = 1$ and
$\lambda =10$, this minimum is 1.5 to almost two times higher than that of
the second chance policy.  For higher query rates, the performance of
the linear and logarithmic policies is less affected by the choice of
$\alpha$.
These results show that choosing a priori an $\alpha$ value for the
linear and logarithmic policies that will perform well across all
workloads is difficult.

The log-based second-chance policy consistently outperforms both
probability-based policies and achieves a total cost very near the
minimum total cost.  This is because, unlike the probability-based
policies that depend on a function of the node's network distance from
the root node, the second-chance policy adapts to the timing of the
queries within the workload and thus accounts for shifts in key
popularity, which is independent of the distance of the node.  In all
remaining experiments, we use second-chance as the cut-off policy.

\subsection{Varying the Network Size}
\label{NetworkSize}

In this section we show that CUP performs well as we vary the size
of the network.

\begin{table*}
\caption[]{\small Comparison of CUP with standard caching for varying
numbers of nodes n = $2^k$ for k between 3 and 12.}
\label{tab:comparison2}
\begin{center}
\begin{tabular}{|l|r|r|r|r|r|r|r|r|r|r|r|} \hline
\textsf{\textbf{Number of Nodes}} & \textsf{\textbf{8}} & \textsf{\textbf{16}} 
  & \textsf{\textbf{32}} & \textsf{\textbf{64}}  & \textsf{\textbf{128}}
  & \textsf{\textbf{256}} & \textsf{\textbf{512}}  & \textsf{\textbf{1024}}
  & \textsf{\textbf{2048}} & \textsf{\textbf{4096}} \\\hline
CUP / STD Caching Miss Cost & 0.47  & 0.41  & 0.36  & 0.20 & 0.19 & 0.23 & 0.17 & 0.15 & 0.15 & 0.15   \\\hline
CUP miss latency &2.3 &2.7 &3.0 &3.0 &3.2 &4.0 & 3.9 & 3.9  & 5.5 & 7.3 \\\hline
STD Caching miss latency &2.8 &3.0 &3.5 &4.4 &5.1 & 5.4 & 7.7 & 9.4 & 13.0 & 19.1 \\\hline
Saved miss hops per CUP overhead hop & 0.77 & 1.05  & 1.47  & 2.70 & 3.44 & 3.0 & 5.57 & 7.05 & 9.13 & 13.2 \\\hline
\end{tabular}
\end{center}
\end{table*}

Table~\ref{tab:comparison2} compares CUP and standard caching for
varying numbers of nodes using three metrics.  We use a $\lambda$ rate
of one query per second.  The first row shows the CUP miss cost as a
fraction of the standard caching miss cost.  The second and third rows
show the query latency measured by average number of hops needed to
handle a miss for CUP and standard caching respectively.  As can be
observed, CUP reduces latency respectively by 5.5, 7.5, and 11.8 hops
per miss for the 1024, 2048, and 4096 node networks.  This is a
substantial reduction (as much as a factor of three) in query response
time in peer-to-peer networks.

The fourth row in Table~\ref{tab:comparison2} shows the ``investment''
return per update push performed by CUP.  This is computed as the
overall ratio of saved miss cost to overhead incurred by CUP:

\small
\[
\label{eq:MissCostSaved}
\begin{split}
{S}&{avedMissOverheadRatio} \\ 
& = \frac {MissCost_{StandardCaching} - MissCost_{CUP}} {OverheadCost_{CUP}}
\end{split}
\] 

\normalsize For the last three network sizes, the return is
respectively 7.05, 9.13, and 13.2 which are remarkable values for a
fully recoverable overhead investment.  The return increases with the
network size, thus CUP is more amenable to larger networks.

Note that this table was generated using a very low query arrival
rate.  As a point of comparison, for a network of 1024 nodes and
\(\lambda = 1000\) queries per second, the CUP miss cost is 0.09 that
of standard caching, the average CUP miss latency, 2.4 hops, is over
ten times less than that for standard caching (25.1 hops), and the
ratio of saved miss cost to update cost is 168 to 1.  This
demonstrates that CUP is indeed amenable to higher query arrival
rates.

\subsection{Multiple Replicas per Key}

In this section we examine the effect on CUP performance of having
multiple replicas per key and propagating updates from each of these
replicas.

A node that is receiving updates for key K will likely have
multiple index entries cached for K.  At first glance it seems
that the more replicas there are in the system, the fewer the
freshness misses for K the node should experience.  

It turns out this will occur only if the cut-off policy is independent
of the number of replicas in the network.  A naive implementation of a
cut-off policy applies its decision at \emph{every} update arrival for
K.  Since the rate of update arrivals for K increases with the number
of replicas in the system, the chances that the node will receive
enough queries to pass its cut-off test are lower, and the naive
implementation mistakenly concludes that there is not enough incentive
to continue receiving updates.  It therefore pushes up a clear-bit
message earlier than necessary.

Table~\ref{tab:replicas} illustrates this problem for a network of 1024
nodes using the second-chance policy and a \(\lambda\) rate of 1 query
per second.  Column 2 in the table shows the miss cost and, in
parentheses, the absolute number of misses incurred by the naive
implementation.  In this case, adding more replicas has the opposite
effect of what we expected.  More replicas can mean more misses.  In
fact, the single replica run exhibits the smallest miss cost and
absolute number of misses.

A solution to this problem requires that the cut-off decision be
independent of the number of replicas in the network.  One way to
achieve this is to trigger the decision only when updates for a
particular replica arrive, and to reset the popularity measure only at
those times.  This ensures that the popularity measure remains the same
across updates for other replicas for the same key.  Column 3 of
Table~\ref{tab:replicas} shows that with this fix, the miss cost and
number of misses do decrease as the number of replicas increases.

\begin{table*}
\caption[]{\small Miss cost, absolute number of misses, and total cost
for varying number of replicas }
\label{tab:replicas}
\begin{center}
\begin{tabular}{|l|r|r|r|} \hline
\textsf{\textbf{ }} & \textsf{\textbf{Naive Cut-Off }} & \textsf{\textbf{Replica-independent Cut-Off}} & \textsf{\textbf{Replica-independent Cut-Off}} \\
\textsf{\textbf{Replicas}} & \textsf{\textbf{Miss Cost \& (Misses)}} 
& \textsf{\textbf{Miss Cost \& (Misses)}} & \textsf{\textbf{Total Cost}} \\\hline
100 & 58068 (4493) & 7562 (502) & 608998 \\ \hline
50 & 59261  (4522) & 7562 (502) & 310301 \\ \hline
10 & 44079  (4296) & 7565 (504) & 69086 \\ \hline
5 & 24406   (2936) & 7565 (504) & 39615 \\ \hline
2 & 11614   (1366) & 7607 (522) & 21050 \\ \hline
1 & 8460    (1206) & 8460 (1206) & 15183  \\ \hline
\end{tabular}
\end{center}
\end{table*}

The last column of Table~\ref{tab:replicas} shows the total cost when
each replica refresh is sent as a separate update.  When compared to
the 55905 hops of total cost for standard caching from
Table~\ref{tab:policies2}, we observe that the total cost of CUP will
eventually overtake that of standard caching as we increase the number
of replicas.  In fact this occurs at eight replicas where the total
cost is 57430.  While these results may seem to imply that a handful
of replicas is enough for good CUP performance, for some applications,
having many more replicas in the network is necessary even if they run
the risk of unrecoverable additional CUP overhead.  For example,
having multiple replicas of content helps to balance the demand for
that content across many nodes and reduces latency.

One may view pushing updates for multiple replicas as an
example of an aggressive CUP policy we refer to in
Section~\ref{CostModel}. At 100 replicas, the total cost is about 10
times that of standard caching.  CUP pays the price of extra overhead
but achieves a miss cost that is about 13.5 percent that of standard
caching.  Therefore, at the cost of extra network load, both query
latency is reduced and the demand for content is balanced across a
greater number of nodes.

If however network load is a concern, there are a couple of techniques an
authority node can use to reduce the overhead of CUP when there are
many replicas in the network. 
First, rather than push all replica refreshes,
the authority node can selectively choose to propagate a subset of the
replica refreshes and suppress others.  This allows the authority node
to reduce update overhead as well as balance demand for content across
the replicas.  Another alternative would be to aggregate replica
refreshes.  When a refresh arrives for one replica, the authority node
waits a threshold amount of time for other updates for the same key to
arrive.  It then batches all updates that arrive within that time and
propagates them together as one update.  This threshold would be a
function of the lifetime of a replica and could be dynamically
adjusted with the number of replicas in the system.  We are
experimenting with different kinds of threshold functions.

\subsection{Varying Outgoing Update Capacity}
Our experiments thus far show that CUP clearly outperforms standard
caching under conditions where all nodes have full outgoing capacity.
A node with full outgoing capacity is a node that can and does
propagate all updates for which there are interested neighbors.  In
reality, an individual node's outgoing capacity will vary with its
workload, network connectivity, and willingness to propagate updates.
In this section we study the effect on CUP performance of reducing the
outgoing update capacity of nodes.

We present two experiments run on a network of 1024 nodes.  In the
first experiment, called "Up-And-Down", after a five minute warm up
period, we randomly select twenty percent of the nodes to reduce their
capacity to a fraction of their full capacity.  These nodes operate at
reduced capacity for ten minutes after which they return to full
capacity.  After another five minutes for stabilization, we randomly
select another set of nodes and reduce their capacity for ten minutes.
We proceed this way for the entire 3000 seconds during which queries
are posted, so capacity loss occurs three times during the simulation.
In the second experiment, called "Once-Down-Always-Down", after the
initial five minute warmup period, the randomly selected nodes reduce
and remain at reduced capacity for the remainder of the experiment.
  
\begin{figure}[tb]
\centerline{\includegraphics[width=9cm]{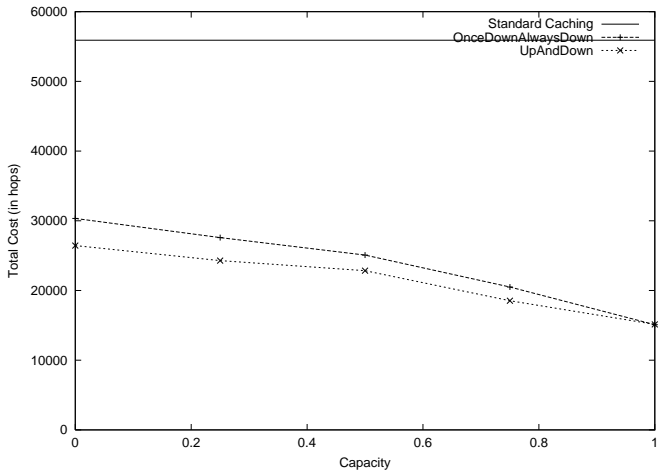}}
\caption{\small Total cost versus reduced capacity.}
\label{fig:capacityQ1}
\end{figure}

\begin{figure}[tb]
\centerline{\includegraphics[width=9cm]{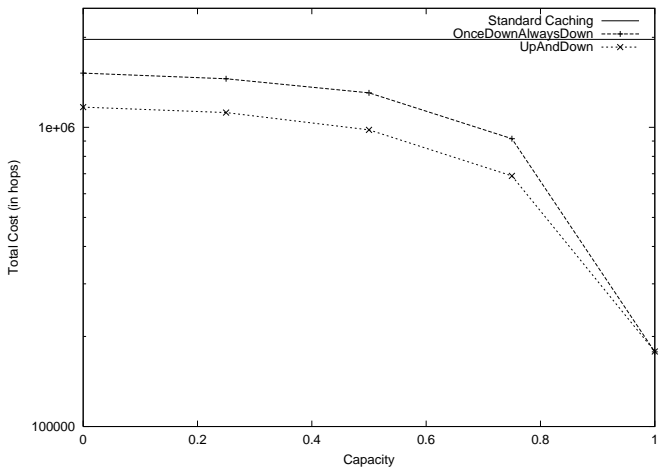}}
\caption{\small Total cost versus reduced capacity. The y-axis is
log scale.}
\label{fig:capacityQ1000}
\end{figure}

Figure~\ref{fig:capacityQ1} shows the total cost incurred by CUP
versus reduced capacity $c$ for both Up-And-Down and
Once-Down-Always-Down configurations.  A reduced capacity $c = .25$
means a node is only pushing out one-fourth the updates it receives.
The figure also shows the total cost for standard caching as a
horizontal line for comparison.  The $\lambda$ rate is 1 query per
second.  Figure \ref{fig:capacityQ1000} shows the same for $\lambda$=
1000 which is especially interesting because CUP has bigger wins with
higher query rates since more updates are justified than with lower
query rates.  Therefore, with high query rates CUP has more to lose if
updates do not get propagated.

Note that even when the capacity of one fifth of the nodes 
is reduced to zero percent  and these nodes do not propagate updates,
CUP outperforms standard
caching for both query rates.  The total cost incurred by CUP is about
half that of standard caching for one  query per second for both
configurations.  For 1000 queries per second, the total cost of CUP is
0.56 and 0.77 that of standard caching for "Up-And-Down" and
"Once-Down-Always-Down" respectively.

A key observation from these experiments is that CUP's performance
degrades gracefully as $c$ decreases.  This is because the reduction
in propagation saves any 
overhead that would have occurred
otherwise.  The important point here is that even if nodes can only
push out a fraction of updates to interested neighbors, CUP still
extends the benefits of standard caching.  Clearly though, CUP
achieves its full potential when all nodes have maximum propagation
capacity.

\section{Related Work}
\label{RelatedWork}

Some peer-to-peer systems suffer from what
we call the ``open-connection'' problem.  Every time a peer node
receives a query for which it does not have an answer cached, it asks
one (e.g., Freenet \cite{clarke00}) or more (e.g., Gnutella \cite{gnutella}) 
neighbors the same query by
opening a connection and forwarding the query through that connection.
The node keeps the connection open until the answer is returned through it.
For every query on every item for which the node does not have a cached
answer, the connection is maintained until the answer comes back.
This results in excessive overhead for the node because it must
maintain the state of many open connections.  CUP avoids this overhead
by asynchronously pushing responses as first-time updates and by
coalescing queries for the same item into one query.

Chord \cite{stoica01} and CFS \cite{dabek01} suggest alternatives to
making the query response travel down the reverse query path back to
the query issuer.  Chord suggests iterative searches where the query
issuer contacts each node on the query path one-by-one for the item of
interest until one of the nodes is found to have the item.  CFS
suggests that the query be forwarded from node to node until a node is
found to have the item.  This node then directly sends the query
response back to the issuer.  Both of these approaches help avoid some
of the long latencies that may occur as the query response traverses
the reverse query path.
CUP is advantageous regardless of whether the
response is delivered directly to the issuer or through the reverse
query path.  However, to make this work for direct response delivery,
CUP must not coalesce queries for the same item at a node into one
query since each query would need to explicitly carry the return
address information of the query issuer. 

All of the above systems (Gnutella, Freenet, Chord, and CFS) enable
caching at the nodes along the query path.  They do not focus on how
to maintain entries once they have been cached.  Cached items are
removed when they expire and refetched on subsequent queries.  For
very popular items this can lead to higher average response time since
subsequent bursts of queries must wait for the response to travel up
and (possibly) down the query path.  CUP can avoid this problem by
refreshing or updating cached items for which there is interest before
they expire.

Consistent hashing work by Karger et al. \cite{karger97} looks at
relieving hot spots at origin web servers by caching at intermediate
caches between client caches and origin servers.  Requests for items
originate at the leaf clients of a conceptual tree and travel up
through intermediate caches toward the origin server at the root of the
tree.  This work uses a model slightly different from the
peer-to-peer model.  Their model and analysis assume
requests are made only at leaf clients and that intermediate caches do
not store an item until it has been requested some threshold number of
times.  Also, this work does not focus on maintaining cache freshness.

Update propagations in CUP form trees very similar to the application-level
multicast trees built by Scribe \cite{rowstron01a}. Scribe
is a publish-subscribe infrastructure built on top
of Pastry \cite{rowstron01b}.  Scribe creates a multicast tree rooted
at the rendez-vous point of each multicast group.
Publishers send a message to the rendez-vous point which then
transmits the message to the entire group by sending it down the
multicast tree.
The multicast tree is formed by joining the Pastry routes from each
subscriber node to the rendez-vous point.  Scribe could apply the
ideas CUP introduces to provide update propagation for cache
maintenance in Pastry.

Cohen and Kaplan study the effect that aging through cascaded caches
has on the miss rates of web client caches \cite{cohen01a}. For each
object an intermediate cache refreshes its copy of the object when its
age exceeds a fraction \emph{v} of the lifetime duration.  The
intermediate cache does not push this refresh to the client; instead,
the client waits until its own copy has expired at which point it
fetches the intermediate cache's copy with the remaining lifetime.
For some sequences of requests at the client cache and some
\emph{v}'s, the client cache can suffer from a higher miss rate than
if the intermediate cache only refreshed on expiration.  Their model
assumes zero communication delay. A CUP tree could be viewed as a
series of cascaded caches in that each node depends on the previous
node in the tree for updates to an index entry.  The key difference is
that in CUP, refreshes are pushed down the entire tree of interested
nodes.  Therefore, barring communication delays, whenever a parent
cache gets a refresh so does the interested child node.  In such
situations, the miss rate at the child node actually improves.

\section{Conclusions}
\label{Conclusions}

In this paper we propose CUP: Controlled Update Propagation for cache
maintenance.  CUP query channels coalesce bursts of queries for the
same item into a single query.  CUP update channels refresh
intermediate caches and reduce the average query latency by over a
factor of ten in favorable conditions, and as much as a factor of
three in unfavorable conditions.  Through light book-keeping, CUP
controls and confines propagations so that only updates that are
likely to be justified are propagated.  In fact, when only half the
number of updates propagated are justified, CUP's overhead is
completely recovered.  Finally, even when a large percentage of nodes
cannot propagate updates (due to limited capacity), CUP continues to
outperform standard caching with expiration.

\section{Acknowledgements}
This research is supported by the Stanford Networking Reseach Center,
and by DARPA (contract N66001-00-C-8015).

The work presented here has benefited greatly from discussions with
Petros Maniatis, Armando Fox, Nick McKeown, and Rajeev Motwani. We
thank them for their invaluable feedback.

\small

\newcommand{\etalchar}[1]{$^{#1}$}


\end{document}